\documentclass[aps,pra,twocolumn,groupedaddress,superscriptaddress]{revtex4-1}
\usepackage{graphicx}
\usepackage{dcolumn}
\usepackage[dvipsnames]{xcolor}
\usepackage{amsmath}
\usepackage{amssymb}
\usepackage{appendix}

\begin{document}

\title{Signatures for strong-field QED in the quantum limit of beamstrahlung}

\author{W. L. Zhang}
\email[]{wenlong.zhang@tecnico.ulisboa.pt}
\affiliation{Engineering Research Center of Nuclear Technology Application, Ministry of Education, East China University of Technology, Nanchang 330013, China}
\affiliation{GoLP/Instituto de Plasmas e Fusão Nuclear, Instituto Superior Técnico, Universidade de Lisboa, Lisboa, Portugal}

\author{T. Grismayer}
\email[]{thomas.grismayer@tecnico.ulisboa.pt}

\author{L. O. Silva}
\email[]{luis.silva@tecnico.ulisboa.pt}
\affiliation{GoLP/Instituto de Plasmas e Fusão Nuclear, Instituto Superior Técnico, Universidade de Lisboa, Lisboa, Portugal}

\date{\today}
\begin{abstract}
{Signatures for strong-field quantum electrodynamics are determined for collisions between round ultrarelativistic leptonic beams in the quantum limit of beamstrahlung. In the low disruption regime, we derive the integrated beamstrahlung photon spectrum that features a characteristic peak close to the beam energy. The conditions to precisely observe this peak experimentally are given regarding the beam parameters. Moreover, the effects of electron-positron pair creation and beam disruption on the photon spectrum are discussed and explored with 3-dimensional particle-in-cell QED simulations. The photon spectrum is associated with the emission of ultrashort and highly collimated gamma-ray beams with a peak spectral brightness exceeding $10^{30}\ \mathrm{photons}/(\mathrm{s}\ \mathrm{mm}^2\ \mathrm{mrad}^2\ 0.1\% \mathrm{BW})$ at $100\ \mathrm{GeV}$-level photon energies (close to the beam energy).}
\end{abstract}


\maketitle

\section{Introduction}

The next-generation high-luminosity linear lepton (electron or positron) colliders \cite{Shiltsev2021,Gray2021,ILC_ExecutiveSummary,CLIC_Implementation2017,FACETII_report,EurStrParPhys2022} can be impacted by beam-beam effects, such as beamstrahlung (synchrotron radiation) and disruption (beam pinch or repulsion) under fields of the oncoming beam \cite{Chen1992,Yokoya1992,Schulte2017}. Beamstrahlung is particularly important for the performance of lepton colliders, as it causes energy loss and degrades the luminosity \cite{Schulte2017,Barklow2023}. For $\mathrm{TeV}$-level colliders \cite{CLIC_Implementation2017}, beamstrahlung becomes significant \cite{Schulte2017,CLIC_Implementation2017}, even with flat beams (specifically designed for minimizing beamstrahlung). The corresponding radiation enters the quantum regime \cite{Schulte2017}, where the particle dynamics is described by the strong-field QED (SF-QED) theory \cite{Ritus1985,Baier1998,Piazza2012,Bulanov2013,Gonoskov2022,Fedotov2022} which predicts  nonlinear Compton scattering (responsible for beamstrahlung explored here) and nonlinear (multi-photon) Breit-Wheeler electron-positron pair creation. The quantum regime is characterized by $\chi _e > 1$, where $\chi _e$ is the quantum parameter defined as $\chi _e=\sqrt{(\gamma {\bf {E}}+ {\bf {p}}/mc\times {\bf {B}})^2-( {\bf {p}}/mc\cdot {\bf {E}})^2}/E_s$. Here, $E_s=m^2c^3/e\hbar$ is the Sauter-Schwinger field, $\hbar$ is the reduced Planck constant, $c$ is the speed of light in vacuum, $e$ and $m$ are the charge and mass of an electron, ${\bf {p}}$ and $\gamma$ are the momentum and the Lorentz factor of the particle, and ${\bf {E}}$ and ${\bf {B}}$ are the electric and magnetic fields.

High-gradient advanced accelerator concepts (AAC) \cite{Schroeder2010,ALEGRO2019,EurStrParPhys2022,Clark2022,Adli2022,England2022,Snowmass21TaskForce} aim to deliver compact (micron- or submicron-scale) and high-energy ($10$'$\mathrm{s\ GeV}$ to a few $\mathrm{TeV}$) leptonic beams with high particle number ($N_0>10^9$ per bunch). With these beams, the colliders will place beamstrahlung in the quantum or even deep quantum ($\chi_{e}\gg 1$) regimes \cite{Noble1987,Chen1989,Chen1992,Yokoya1992,Fabrizio2019,Yakimenko2019,Tamburini2020,Samsonov2021}. This beam-beam scenario presents a controlled and clean platform for probing the SF-QED signatures, as an alternative to the laser-based configurations \cite{Burke1997,Blackburn2014,Piazza2016,Vranic2016,Lobet2017,Gales2018,Poder2018,Niel2018b,Cole2018,Baumann2019b,Albert2020,Hu2020,Fedeli2021,Grech2021,Qu2021,Golub2022,Turner2022,Ahmadiniaz2022}, beam-crystal interaction \cite{Piazza2020}, and beam-plasma interaction \cite{Matheron2022}. AAC-based colliders may intrinsically deliver round or nearly round beams \cite{Fabrizio2019,Yakimenko2019,Barklow2023}. Beamstrahlung and disruption are augmented with round beams, further motivating the study of quantum-dominated beam-beam effects to prepare for future lepton colliders \cite{Barklow2023}.

Previously, Del Gaudio \emph{et al.} \cite{Fabrizio2019} studied beamstrahlung and pair creation with round colliding beams in the weak quantum regime ($\chi_e \lesssim 1$). The commonly used mean field model \cite{Noble1987,Chen1992}, with the realistic field profiles discarded, was shown to deviate from the theoretical beamstrahlung spectrum \cite{Fabrizio2019}. Recently, Tamburini \emph{et al}. \cite{Tamburini2020} proposed a specially designed, asymmetric collision configuration where the single-particle radiation dynamics can be investigated. However, in realistic (either laser- or beam-driven) experiments using leptonic beams with round or nearly round shape, the observed SF-QED signatures, e.g., photons collected by the diagnostics, will include the integrated contribution from all particles with different $\chi_e$ in the beam(s). This integrated photon emission and its dependence on the associated beam-beam effects remain to be explored.

In this article, we study beamstrahlung in the quantum regime ($\chi_e \gtrsim 1$), and we identify a clear signature for SF-QED as a sharp peak close to the beam energy in the photon spectrum. This distinctive signature is found to appear at high $\chi_e$ (above a threshold value) and be susceptible to the impact of beam-beam effects and energy spread of the colliding beams. Yakimenko \emph{et al}. \cite{Yakimenko2019} have shown that the collision of tightly compressed beams can enable a significant fraction of the beam particles to enter the fully nonperturbative QED regime \cite{Ritus1972, Fedotov2017,Podszus2019,Ilderton2019_NPQED,Mironov2020,Heinzl2021}, i.e., $\alpha \chi_e^{2/3}\gtrsim 1$ ($\chi_e\gtrsim 1600$) with $\alpha\simeq 1/137$ being the fine-structure constant. The exact theory for this nonperturbative regime has not been established. Here, our work focuses on the quantum regime, but with $\alpha \chi_e^{2/3}\ll 1$, allowing us to explore the signatures of SF-QED in the perturbative regime.

This paper is organized as follows. An analytical solution to the integrated beamstrahlung spectrum in the $\chi_{e}\gtrsim 1$ regime is given in Sec. \ref{sec: analytical_spectrum}. Section \ref{sec: 3D_PIC_simulation} describes the particle-in-cell simulations that support our study. Section \ref{sec: discussions} is devoted to the beamstrahlung properties and their dependence on the beam parameters, impact of beam-beam effects (pair creation and beam disruption), and misalignment (beam offset). The validity of the locally constant field approximation (LCFA) in leptonic collisions is thoroughly examined in Sec. \ref{sec: LCFA}. Section \ref{sec: quality_gammaray} characterizes the quality of the emitted gamma-ray photon beams in leptonic collisions. Especially, the brightness of the gamma-ray beam is analyzed and theoretically estimated. The conclusions of this paper are summarized in Sec. \ref{sec: conclusion}. Appendix \ref{sec: appendix_Analytical_spectrum} and \ref{sec: appendix_Brightness} provide the derivation details of this paper.

\section{Integrated beamstrahlung spectrum in the quantum regime}
\label{sec: analytical_spectrum}

An analytical expression for the integrated beamstrahlung spectrum can be derived when pair creation and the disruption effects are negligible, i.e., $D \ll 1$, where $D=r_eN_0\sigma_z/\gamma \sigma_0^2$ is the disruption parameter \cite{Chen1988}. Here, $r_e=e^2/mc^2$ is the classical electron radius, $\sigma_z$ and $\sigma_0$ are the longitudinal length and transverse size of the beam, respectively. After the integrated beamstrahlung spectrum is derived, the impact of these beam-beam effects (pair creation and beam disruption) will be discussed and demonstrated by particle-in-cell simulations in Sec. \ref{sec: impact_beambeam_effects}.

\subsection{Beam parameters and setup}
\label{sec: parameter_setup}

We consider collisions between two beams with Gaussian density profiles characterized by $n=n_0 f(r) g(z)$. Here, $n_0$ is the peak density, $f(r)=\exp (-r^2/2\sigma_{0}^2)$, where $r$ is the radial position, assuming round (cylindrical) symmetry in the transverse direction. $g(z)=\exp ( -(z-z_c)^2/2\sigma_z^2 )$, where $z$ is the longitudinal coordinate co-moving with the beam and $z_c$ depicts the beam center. The radial electric and azimuthal magnetic fields of an ultrarelativistic beam are given by $E_r=4\pi e n_0 F(r) g(z)$ and $B_\theta \simeq E_r$ \cite{Fabrizio2019}, with $F(r)=\int_{0}^{r} f(r')r'dr'/r$. The local, instantaneous $\chi_e$ for a particle in one beam is then given by
\begin{equation}
\chi_e (r,t)\simeq \frac{2\gamma E_r}{E_s}=\frac{8\pi |e|n_0\gamma }{E_s}F(r)\exp (-u^2),
\label{Eq: chi_e}
\end{equation}
where $u=\sqrt{2} c t/\sigma_z$ is the normalized time, and we have assumed that the particle under study crosses the center of the oncoming beam at $t=0$. For a Gaussian beam,  $F(r)=\sigma_0^2(1-\exp(-r^2/2\sigma_0^2))/r$, and the fields peak at $r=r_{peak}=1.6 \, \sigma_0$, leading to the maximum $\chi_{e}$ given by 
\begin{equation}
    \chi_{e \, max}=\chi_{e}(r_{peak}, 0)=15.3\frac{\mathcal{E}_0[\mathrm{10 \, GeV}]\  N_0[10^{10}]}{\sigma_0[0.1 \mu m]  \ \sigma_z[0.1 \mu m]},
    \label{Eq: chi_emax}
\end{equation}
where $\mathcal{E}_0=\gamma m c^2$ is the beam energy. Equation \eqref{Eq: chi_emax} shows that the $\chi_{e}> 1$ regime can be reached by delivering high-energy ($\mathcal{E}_0>10\ \mathrm{GeV}$) and submicron ($\sigma_0\lesssim \mu m$ and $\sigma_z\lesssim \mu m$) beams with $\sim \mathrm{nC}$ charge.

\subsection{Integrated photon spectrum}

The main steps of the derivation of the integrated photon spectrum are given below but additional technical details can be found in Appendix \ref{sec: appendix_Analytical_spectrum}.
The differential probability rate for single photon emission in the quantum regime is given by \cite{Ritus1985}
\begin{equation}
\frac{d^2 W}{dt d\xi}=\frac{\alpha}{\sqrt{3}\pi \tau_c \gamma }\left[\mathrm{Int}K_{5/3}(b)+ \frac{\xi ^2}{1-\xi}K_{2/3}(b) \right],
\label{Eq: W_omega}
\end{equation}
where $\tau_c = \hbar/mc^2$ is the Compton time, $\xi = \mathcal{E}_\gamma /\mathcal{E}_0$ with $\mathcal{E}_\gamma$ the photon energy, $b=2/(3\chi_{e}(r,t))\xi/(1-\xi)$ and $\mathrm{Int}K_{5/3}(b)=\int_{b}^{\infty}K_{5/3}(q)dq$, with $K_{\nu}$ the modified Bessel function of the second kind. Equation \eqref{Eq: W_omega} is established with the locally constant field approximation (LCFA) \cite{Ritus1985,Piazza2018,Piazza2019,Ilderton_PRA_LCFA_2019,King2020,Lv2021,Gonoskov2022,Fedotov2022}. The dependence of the validity of LCFA on collision parameters and photon energies is analyzed in Sec. \ref{sec: LCFA}. We have verified that the LCFA can be readily employed for the collisions considered in our study. The time-integrated photon spectrum emitted from a single particle in a beam is given by \cite{Fabrizio2019}
\begin{equation}
s_\omega (\xi,r)=\int_{-\infty}^{\infty}\frac{d^2 W}{dt d\xi} dt. 
\end{equation}
The integrated radiation spectrum from the beam, normalized by the particle number $N_0$, is then given by 
\begin{equation}
\mathcal{S}_\omega(\xi)=\int_0^{\infty} s_\omega (\xi,r)f(r)rdr\left(\int_0^{\infty}f(r)rdr\right)^{-1}.  
\end{equation}
In order to derive $\mathcal{S}_\omega(\xi)$, we first approximate the special functions in $d^2W/dtd\xi$ [Eq. \eqref{Eq: W_omega}], such that $K_{2/3}(b) \simeq k_{2/3}b^{-2/3}\exp (-b)$ and $\mathrm{Int}K_{5/3}\simeq 2K_{2/3}(b)\simeq 2k_{2/3}b^{-2/3}\exp (-b)$, where $k_{2/3}=1.23$ is a fitting coefficient. With these expressions, $d^2W/dtd\xi$ can be well approximated. The details of this approximation and the corresponding numerical verification are given in Appendix \ref{sec: appendix_Analytical_spectrum}. 

In addition, we approximate the density profile of the beams as a uniform profile. Since the radiation probability [Eq. \eqref{Eq: W_omega}] is solely determined by $\chi_e(r,t)$ of the emitting particle, it is crucial to preserve the $\chi_e$ distribution of particles in the collision ($dN/d\chi_e$) with the approximated profiles. For this purpose, we approximate the Gaussian profiles by a cylinder of radius $\sigma _{0A}$ and length $\sigma_{zA}=2\sqrt{2}\sigma_z$, with a constant density $n_{0A}$. We also require this approximated profile to match $\chi_{e \, max}$ of the original Gaussian profiles, leading to $n_{0A}\sigma_{0A}=0.9n_0\sigma_0$. The detail for characterizing the approximated profile is given in Appendix \ref{sec: appendix_Analytical_spectrum}.
 \begin{figure}
	\includegraphics[width=8.5cm,height=5.43cm]{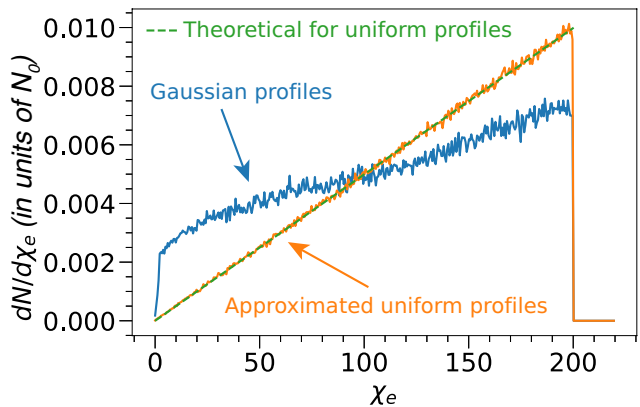}
	\caption{(Color online). The $dN/d\chi_e$ distribution of particles in leptonic beam-beam collisions. A Monte Carlo code is developed, where the colliding beams are represented by pseudo-particles created using random number generation. Each pseudo-particle is assigned with a weight given by the corresponding initial local density within the beam. The beams are set to cross each other at the speed of light. The $\chi_e(r,t)$ of each particle is calculated based on the local fields from the oncoming beam. The $dN/d\chi_e$ distribution shown here is taken at the peak interaction moment (when the centers of colliding beams cross) in a collision with $\chi_{e \, max}=200$. The distribution is normalized by the particle number of the beam ($N_0$). $10^6$ pseudo-particles are initialized for each colliding beam. Different beam profiles are compared here. The blue line is for Gaussian colliding beams with energy $\mathcal{E}_0=90\ \mathrm{GeV}$, longitudinal length $\sigma_{z}=30.7\ \mathrm{nm}$, transverse size $\sigma_{0}=30.7\ \mathrm{nm}$, and particle number $N_0=1.37\times 10^9$. The orange line is for the approximated uniform profile as specified in the text. The green dashed line represents the theoretical distribution for the approximated uniform profile, i.e., $dN/d\chi_e=2\chi_e/\chi_{e \, max}^2$.}
	\label{fig: Gaus_Unif_justification}
\end{figure}

Here, we justify that a collision with the approximated uniform profile introduced above can preserve the $dN/d\chi_e$ distribution of particles in the collision between the original Gaussian beams. Figure \ref{fig: Gaus_Unif_justification} depicts the $dN/d\chi_e$ distribution at the peak interaction moment (when the centers of the two beams cross) in a collision with $\chi_{e \, max}=200$. We have developed a Monte Carlo code to simulate beam-beam collisions where the $dN/d\chi_e$ distribution can be conveniently calculated, as explained in the figure caption of Fig. \ref{fig: Gaus_Unif_justification}. The theoretical $dN/d\chi_e$ of the approximated uniform profile can be analytically given, i.e., $dN/d\chi_e=2\chi_e/\chi_{e \, max}^2$ (green line), and the Monte Carlo result (orange line) perfectly agrees with the analytical result. In addition, the approximated uniform profile is shown to largely preserve the shape and amplitude of $dN/d\chi_e$ distribution, compared with those of original Gaussian profiles (blue line). This preservation assures that the quantum radiation calculated with the approximated uniform profile can well reflect the radiation from collisions between Gaussian beams.

With the above approximations, $\mathcal{S}_\omega(\xi)$ is given by
\begin{equation}
\mathcal{S}_\omega(\xi) = \frac{2\sqrt{2}\alpha \sigma_{z}}{\sqrt{3}\pi\tau_c\gamma c} k_{2/3} \left(2+\frac{\xi ^2}{1-\xi} \right) b_0^2 \ \Gamma \left[-\frac{8}{3}, b_0\right],
\label{Eq: S_omega}
\end{equation}
where $b_0=2/(3\chi_{e \, max})\xi/(1-\xi)$, and $\Gamma$ is the incomplete Gamma function \cite{IncGamma_Book,IncGamma_Online}. The detailed derivation is provided in Appendix \ref{sec: appendix_Analytical_spectrum}. The analytical spectrum Eq. \eqref{Eq: S_omega} is verified to give the accurate photon emission in both the $\chi_{e \, max}\sim 1$ and the $\chi_{e \, max}> 1$ regimes. We found that the mean field model \cite{Chen1992} severely underestimates the photon emission in the $\chi_{e \, max}> 1$ regime, because this model assumes a low mean $\chi_e$, i.e., $\chi_{e \, \mathrm{mean}}\simeq 0.6\chi_{e \, max}$, for all the particles. This average treatment will suppress the critical contribution from particles with high-$\chi_e$ ($\sim \chi_{e \, max}$).

\section{Three-dimensional particle-in-cell simulations}
\label{sec: 3D_PIC_simulation}

\begin{figure}
	\includegraphics[width=8.5cm, height=5.37cm]{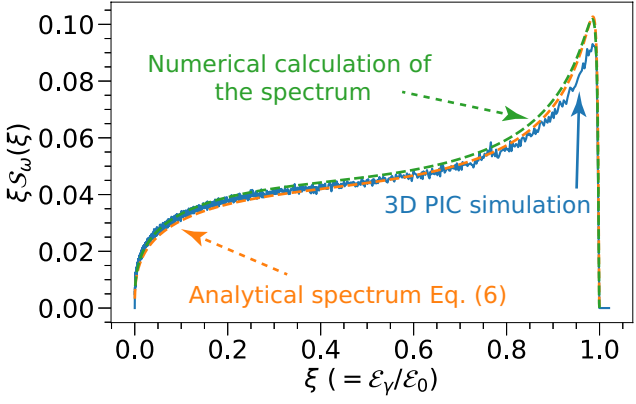}
	\caption{(Color online). The normalized energy spectrum $\xi \mathcal{S}_\omega (\xi)$ of photon radiation from a head-on collision between two identical, cold, round, and Gaussian electron beams with $\chi_{e \, max}=200$. The beam parameters are the same with those used in Fig. \ref{fig: Gaus_Unif_justification}, i.e., $\mathcal{E}_0=90\ \mathrm{GeV}$, $\sigma_0=30.7 \ \mathrm{nm}$, $\sigma_z=30.7\ \mathrm{nm}$, and $N_0=1.37\times 10^9$. The blue line is for a 3D PIC simulation. The orange dashed line is for the analytical spectrum Eq. \eqref{Eq: S_omega}. The green dashed line is for the direct numerical calculation of $\mathcal{S}_\omega (\xi)$ using the original Gaussian beam profile.}
	\label{fig: Sw_Symp_chi200}
\end{figure}

3D QED-PIC simulations with {\small{OSIRIS}} \cite{Fonseca2002}, where SF-QED effects are self-consistently included \cite{Grismayer2016,Vranic2016,Zhang2021,Grismayer2021}, have been performed to study the leptonic beam-beam collisions. A typical energy spectrum $\xi \mathcal{S}_\omega (\xi)$ from a head-on collision is shown in Fig. \ref{fig: Sw_Symp_chi200} for an electron-electron collision with $\chi_{e \, max}=200$. In the simulation, the colliding beams counter-propagate along the $z$ direction, with their densities initialized as $n_1=n_0\exp (-(x^2+y^2)/2\sigma_{0}^2)\exp ( -(z+3\sigma _z)^2/2\sigma_z^2)$ for $-6\sigma_z \le z \le 0$, and $n_2=n_0\exp (-(x^2+y^2)/2\sigma_{0}^2)\exp ( -(z-3\sigma _z)^2/2\sigma_z^2)$ for $0\le z \le 6\sigma_z$, respectively, where $n_0=3\times 10^{24}\ \mathrm{cm^{-3}}$. The beams are transversely truncated at $3\sigma_0$, i.e., $\sqrt{x^2+y^2}\le 3\sigma_0$. The computation cells are given by $dz=0.6\ \mathrm{nm}=0.02\sigma_z$ and $dx=dy=0.46\ \mathrm{nm}=0.015\sigma_0$. The time step is $dt=0.009\sigma_z/c$. The number of macro particles of each beam is $7.7\times 10^7$. The simulation lasts until $t=6\sigma_z/c$ when the beams completely cross each other.

For these conditions, the disruption effects are negligible, as $D= 7\times 10^{-4}\ll 1$. The analytical spectrum Eq. \eqref{Eq: S_omega} (orange dashed line in Fig. \ref{fig: Sw_Symp_chi200}) is in excellent agreement with the direct numerical calculation of $\mathcal{S}_\omega(\xi)$ without approximations, i.e., using the exact $d^2W/dtd\xi$ (given by Eq. \eqref{Eq: W_omega}) and the Gaussian profiles. The PIC simulation also confirms our analytical results. For this particular scenario, only $3\%$ of the photon energy is transferred to pairs, indicating that pair creation is negligible, as required by our model. The beam loss (beam-to-photon conversion efficiency) is shown to be $\eta_\gamma = 4.8\%$, in excellent agreement with the theoretical value $\eta_\gamma =\int_{0}^{1}\xi\mathcal{S}_\omega(\xi)d\xi$ using Eq. \eqref{Eq: S_omega}. The theoretical $\eta_\gamma$ will be given explicitly later (using Eq. \eqref{Eq: E_gamma_average}).

\section{Beamstrahlung properties}
\label{sec: discussions}

\subsection{Spectral peak}

Using Eq. \eqref{Eq: S_omega}, we can identify the transition of the beamstrahlung dynamics from the mild quantum regime ($\chi_e \sim 1$) to the deep SF-QED regime ($\chi_e \gg 1$); this transition is found to be characterized by a threshold quantum parameter $\chi_\mathrm{thr} = 40$. When $1\lesssim \chi_{e \, max}<\chi_\mathrm{thr}$, the yield of high-energy ($\xi \sim 1$) photons grows with $\chi_{e \, max}$, forming a plateau-like high-energy tail in the photon spectrum. The energy spectrum ($\xi \mathcal{S}_\omega (\xi)$) is also broadband. When $\chi_{e \, max}>\chi_\mathrm{thr}$, a distinctive, sharp peak close to the beam energy appears in the $\mathcal{S}_\omega(\xi)$ spectrum, as well as in the $\xi \mathcal{S}_\omega (\xi)$ spectrum as shown in Fig. \ref{fig: Sw_Symp_chi200}, where the peak is at $\xi _{peak}=0.985$. This peak shows that the integrated beamstrahlung spectrum from a collision preserves the characteristic peak of the single-particle spectrum, resulting from $d^2W/dtd\xi$ [Eq. \eqref{Eq: W_omega}] for particles with $\chi_e>16$, and first identified in \cite{Esberg2009,Bulanov2013,Tamburini2020}. The increased threshold $\chi_\mathrm{thr}$ is due to the realistic temporal and spatial beam profiles. Physically, a fraction of the beam particles samples a $\chi_e(r,t)$ [Eq. \eqref{Eq: chi_e}] below the threshold $\chi_{e}(=16)$ predicted for observing the peak in the single-particle spectrum, leading to the higher threshold $\chi_\mathrm{thr}=40$.

The characteristic spectral peak can be determined from Eq. \eqref{Eq: S_omega} by solving $\partial_{b_0} (\xi \mathcal{S}_\omega(\xi)) = 0$. The peak position $b_{0,peak}=2/(3\chi_{e \, max})\xi_{peak}/(1-\xi _{peak})$ converges to an asymptotic value, i.e., $b_{0,peak}\rightarrow 0.229$, when $\chi_{e \, max}$ increases. Assuming $b_{0,peak}\simeq 0.229$, Eq. \eqref{Eq: S_omega} determines the amplitude of the spectral peak, $P_{\xi \mathcal{S}_\omega}$, which scales as
\begin{equation}
    P_{\xi \mathcal{S}_\omega} = 0.155\frac{\alpha \sigma_z}{\tau_c\gamma c}\chi_{e \, max} = 0.15\frac{\sigma_{z}[0.1\mathrm{\mu m}]}{\mathcal{E}_0[\mathrm{GeV}]}\chi_{e \, max}.
    \label{Eq: P_xiSw_Sym}
\end{equation}
Since $\xi_{peak}\simeq 1$, $P_{\xi \mathcal{S}_\omega}$ [Eq. \eqref{Eq: P_xiSw_Sym}] can also be used to evaluate the amplitude of $\mathcal{S}_\omega$ spectral peak, $P_{ \mathcal{S}_\omega}$, i.e., $P_{ \mathcal{S}_\omega}\simeq P_{\xi \mathcal{S}_\omega}$. 

\subsection{Average photon number and photon energy}

\begin{figure}
	\includegraphics[width=8.5cm, height=6.01cm]{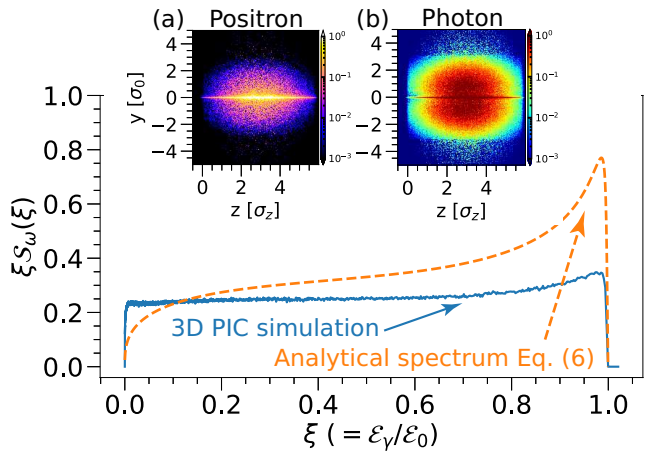}
	\caption{(Color online). The energy spectrum of photons from the same electron-electron collision in Fig. \ref{fig: Sw_Symp_chi200}, but with stronger beamstrahlung by employing longer beams. Here, $\sigma_z=230\ \mathrm{nm}$, leading to $\overline{N_\gamma}=1.5$ according to our theory (using Eq. \eqref{Eq: N_gamma_average}). The initial disruption is weak with $D_0=0.04$. The other parameters, including $\mathcal{E}_0$, $\sigma_{0}$, $n_0$, and $\chi_{e \, max}$, remain unchanged. Insets: density maps at the $y-z$ slice across the beam axis, at the end of the collision. (a) positrons from the created pairs in one colliding beam; (b) corresponding photons radiated from the same beam.}
	\label{fig: N_gamma_1d5}
\end{figure}

The theoretical results (Eq. \eqref{Eq: S_omega} and Eq. \eqref{Eq: P_xiSw_Sym}) can accurately characterize beamstrahlung when beam particles emit less than one photon on average. This requirement can be formulated as $\overline{N_\gamma}<1$, where $\overline{N_\gamma}=N_\gamma /N_0=\int_0^{1}\mathcal{S}_\omega(\xi)d\xi$ is the average number of photon emission from a beam particle, with $N_\gamma$ the number of photons. If strong beamstrahlung occurs ($\overline{N_\gamma}\gtrsim 1$), multiphoton emission becomes significant, and the reported characteristic spectral peak will be smeared, to be demonstrated later (in Fig. \ref{fig: N_gamma_1d5}). Using Eq. \eqref{Eq: S_omega}, in the limit of $\chi_{e \, max}\gg 1$, $\overline{N_\gamma}$ is given by $\overline{N_{\gamma}}\simeq 1.88 \alpha \sigma_{z}\chi_{e \, max}^{2/3}/\tau_c\gamma c$. The condition $\overline{N_\gamma} < 1$ then determines the required collision parameters as
\begin{equation}
\overline{N_{\gamma}} \simeq 1.82 \frac{\sigma_{z}[0.1\mathrm{\mu m}]}
 {\mathcal{E}_0[\mathrm{GeV}]}\chi_{e \, max}^{2/3} < 1 .
\label{Eq: N_gamma_average}
 \end{equation}

The average rate of photon emission from a beam particle in the collision process is $d\overline{N_\gamma}/dt \simeq \overline{N_\gamma}/(\sqrt{2}\sigma_z/c) = 1.33\alpha/(\tau_c\gamma)\chi_{e \, max}^{2/3} $. The scaling of $d\overline{N_\gamma}/dt$ here is identical to the instantaneous rate of photon emission from a particle in the $\chi_e\gg 1$ limit \cite{Gonoskov2022}. This shows that the integrated beamstrahlung dynamics of a collision is characterized by the corresponding $\chi_{e \, max}$, and shares the same scaling as with the single-particle dynamics. Similarly, using Eq. \eqref{Eq: S_omega}, the average energy $\overline{\mathcal{E}_\gamma}$ radiated from a beam particle, $\overline{\mathcal{E}_\gamma}=\int _0^{1}\mathcal{E}_0\xi\mathcal{S}_\omega(\xi)d\xi$, can be obtained as
\begin{equation}
    \overline{\mathcal{E}_\gamma}\simeq 0.47\frac{\alpha mc^2\sigma_{z}}{\tau_c c}\chi_{e \, max}^{2/3} \simeq \frac{\overline{N_\gamma}}{4}\mathcal{E}_0 .
    \label{Eq: E_gamma_average}
\end{equation}
Using Eq. \eqref{Eq: E_gamma_average}, the beam loss is given by $\eta_\gamma=\overline{\mathcal{E}_\gamma}/\mathcal{E}_0\simeq \overline{N_\gamma}/4$, indicating a low beam loss when $\overline{N_\gamma} < 1$. The average radiation power from a beam particle is then $\overline{\mathrm{P_{rad}}}\simeq \overline{\mathcal{E}_\gamma}/(\sqrt{2}\sigma_z/c) = 0.33(\alpha mc^2/\tau_c)\chi_{e \, max}^{2/3}$. We note that the total radiated energy is $\mathcal{E}_\gamma ^{tot}=N_0\overline{\mathcal{E}_\gamma} \propto (\mathcal{E}_0\sqrt{\sigma_z}/\sigma_0)^{2/3}N_0^{5/3}$, showing a strong dependence with the particle number as $\propto N_0^{5/3}$.

We have also examined the effects of finite beam emittance and energy spread on beamstrahlung. The transverse emittance has negligible impact on the particle trajectory and photon emission for collisions with short beams. PIC simulations show that the photon spectrum in Fig. \ref{fig: Sw_Symp_chi200} remains unchanged when the beams are set to have a significant normalized emittance up to $\varepsilon_N\sim 27000\ \mathrm{\mu m}$ (with the divergence angle $\sim 5 \ \mathrm{mrad}$). This emittance is comparable to $\varepsilon_N$ for the self-injected beams from laser wakefield acceleration at the 10 GeV or higher energy scales (see \cite{Martins2010}). The energy spread $\Delta \mathcal{E}_0$ of the colliding beams is found to mainly affect (broaden) the characteristic spectral peak, because $\Delta \mathcal{E}_0$ results in a spread in $\chi_{e \, max}$ according to Eq. \eqref{Eq: chi_emax}. We have verified by PIC simulations that the well-defined spectral peak can be observed for relatively large energy spreads $\Delta \mathcal{E}_0/\mathcal{E}_0<5\%$ which is also achievable according to \cite{Martins2010}.

Our study indicates that the expected signature for SF-QED as the spectral peak can be observed with collision parameters satisfying
\begin{subequations}
	\begin{align}
& \frac{\sigma_z[0.1 \mu m]}{\mathcal{E}_0[\mathrm{GeV}]} < 0.038 \frac{N_0[10^{10}]}{\sigma_0[0.1 \mu m]}\  (\mathrm{for}\ \chi_{e \, max}>\chi_\mathrm{thr}) \label{subeqn: 1} \\	
& \frac{\sigma_z[0.1 \mu m]}{\mathcal{E}_0[\mathrm{GeV}]}<0.07\bigg( \frac{\sigma_0[0.1 \mu m]}{N_0[10^{10}]} \bigg)^2,\ \Delta \mathcal{E}_0/\mathcal{E}_0<5\%, \label{subeqn: 2}
	\end{align}
 \label{Eq: requirement_beam_SFQED}
\end{subequations}
where $\chi_\mathrm{thr}=40$ (analyzed before), and Eq. \eqref{subeqn: 2} comes from the condition $\overline{N_\gamma}<1$ [Eq. \eqref{Eq: N_gamma_average}] and the requirement of energy spread, respectively. For example, for beams with $\mathcal{E}_0=30\ \mathrm{GeV}$ and $\sim \mathrm{nC}$ charge, Eq. \eqref{Eq: requirement_beam_SFQED} indicates that submicron beams are preferred. For $\sigma_0=0.05\ \mu m$, then this condition yields $\sigma_z<0.13\ \mu m$. 

\subsection{The interplay between beamstrahlung and disruption}
\label{sec: impact_beambeam_effects}

The spectral peak is less prominent as pair creation becomes more important. The average number of generated pairs in a collision, given by $\overline{N_p}=N_p/N_0$ where $N_p$ is the number of pairs, can be estimated as follows. The rate of pair creation is approximated by $d\overline{N_p}/dt\simeq 0.38\alpha \chi_{e \, max}^{2/3}/\tau_c\gamma$ in the $\chi_e\gg 1$ regime \cite{Gonoskov2022,Schoeffler2019}. $N_p$ is then $N_p\sim N_{\gamma}(d\overline{N_p}/dt)(\sqrt{2}\sigma_z/c)\simeq 0.28\overline{N_{\gamma}}^2N_0$. Since $\overline{N_p}\propto \overline{N_{\gamma}}^2$, pair creation is negligible when $\overline{N_{\gamma}}<1$. If $\overline{N_{\gamma}}\gtrsim 1$, pair creation will become significant and needs to be taken into account e.g., for designing the collider diagnostics. It is important to stress that the configuration proposed here, with dense, short beams, provides strong and localized fields, ensuring access to the SF-QED regime, and considerable beamstrahlung, but with a negligible pair production that would be deleterious to observing the SF-QED signatures.

Another important effect to discuss is beam disruption. This effect can be related to beamstrahlung via $\overline{N_\gamma}\simeq 1.51\alpha^{4/3}N_0^{1/3}D^{1/3}$. For conventional $\sim \mathrm{nC}$ beams ($N_0\sim 6\times 10^9$), $\overline{N_\gamma}\sim 3.9D^{1/3}$. If $\overline{N_\gamma}<1$ is satisfied, $D\ll 1$, indicating that for our conditions disruption can be safely neglected.

We have studied the mild-disruption ($D\sim 1$), weak beamstrahlung ($\overline{N_\gamma}<1$) regime, in order to clearly show the impact of disruption on beamstrahlung in a decoupled way. This regime can be accessed using transversely compressed ($\sigma_0\lesssim \mathrm{nm}$), $\sim \mathrm{pC}$ ($N_0\sim 10^7$) beams. A 3D PIC simulation for an electron-electron collision, where $\overline{N_\gamma}\simeq 0.5$, $D=1$, and $\chi_{e \, max}=100$, was performed using beams with $\mathcal{E}_0=90\ \mathrm{GeV}$, $N_0=1.28\times 10^{7}$, $\sigma_{0}=0.15\ \mathrm{nm}$, and $\sigma_{z}=115\ \mathrm{nm}$. Due to the disruption effect, the beams expel each other, leading to the drop in beam densities and fields. This mild disruption does not fundamentally change our predictions. The simulation suggests that the emitted photon spectrum approximately agrees with our model [Eq. \eqref{Eq: S_omega}], with a reduced characteristic spectral peak. The photon energy (or beam loss) is reduced by $\lesssim 20\%$ compared to our theory [Eq. \eqref{Eq: E_gamma_average}]. 

The coupled regime between beam disruption and individual stochastic SF-QED events occurs with strong beamstrahlung ($\overline{N_\gamma}\gtrsim 1$) and finite disruption effects. We investigated this regime with 3D PIC simulations, as shown in Fig. \ref{fig: N_gamma_1d5} for the same electron-electron collision as in Fig. \ref{fig: Sw_Symp_chi200}, but using longer beams to access the stronger beamstrahlung regime; this resulted in $\overline{N_{\gamma}}=1.5$ with initial disruption $D_0 = 0.04$. The beam energy loss is $\eta_\gamma \simeq 32\%$, which is still in good agreement with Eq. \eqref{Eq: E_gamma_average}. Pair production becomes significant enough $\overline{N_p}\simeq 0.23$, taking away $\sim 20\%$ of the photon energy, and leading to the reduced emission spectrum (blue line in Fig. \ref{fig: N_gamma_1d5}) as compared to our theory (orange line). The multiphoton emission from beam particles and the radiation from the created pairs result in a considerable number of low-energy photons, forming the lifted low-$\xi$ tail in the spectrum (blue line in Fig. \ref{fig: N_gamma_1d5}). Whereas disruption is initially small, it will dynamically increase due to the beamstrahlung-induced beam loss \cite{Samsonov2021}. The secondary pairs, created at lower energies, are particularly susceptible to disruption effects (as $D\propto \gamma^{-1}$). This is shown in Fig. \ref{fig: N_gamma_1d5}(a), where the created positrons are attracted towards the axis by fields of the oncoming beam, forming a bright density filament on axis. There is also a dense photon filament close to the axis, as shown in Fig. \ref{fig: N_gamma_1d5}(b) -- these photons are radiated from the pairs accumulated there. These exotic phenomena present a probe for disruption-affected SF-QED regimes in beam-beam collisions and future lepton colliders. The regime, where $\overline{N_\gamma}>1$ and $D_0\gtrsim 1$, features strong coupling between disruption and SF-QED effects \cite{Samsonov2021}, and it will be explored in future works.

\begin{figure*}
	\includegraphics[width=10cm,height=7.04cm]{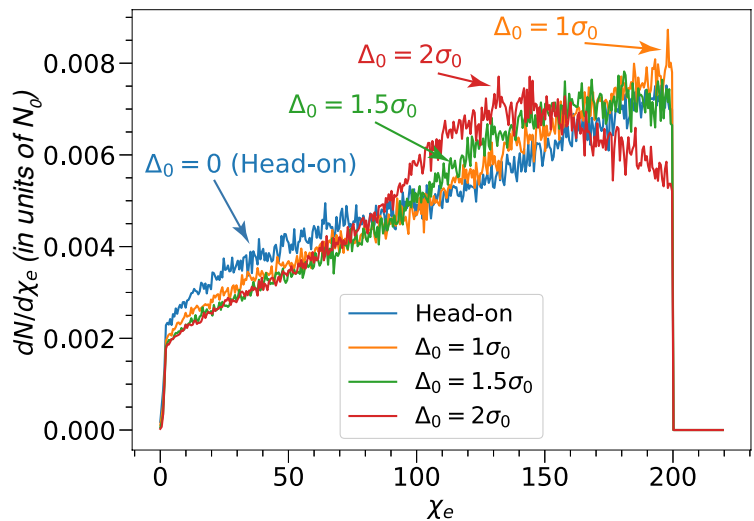}
	\caption{(Color online). The $dN/d\chi_e$ distribution of particles in the same leptonic collision but with different misalignment $\Delta_0$ (offset between the colliding beams). The beam parameters are the same with those used in Fig. \ref{fig: Gaus_Unif_justification} and Fig. \ref{fig: Sw_Symp_chi200}, and the distribution is obtained using the same Monte Carlo method with Fig. \ref{fig: Gaus_Unif_justification}.} 
	\label{fig: misalignment_D0}
\end{figure*}

\subsection{Impact of misalignment on beamstrahlung}
\label{sec: misalignment}

The radial misalignment between the two beams is generally a problem for the particle colliders \cite{ILC_TDR}. Here, we show that collisions with round beams have a high tolerance of offset (set to be $\Delta_0$). Figure \ref{fig: misalignment_D0} demonstrates that the $dN/d\chi_e$ distribution of particles vary slowly with $\Delta_0$, up to $\Delta_0 \gtrsim r_{peak}(=1.6\sigma_0)$. With this robust $dN/d\chi_e$ distribution, the beamstrahlung spectrum will be insensitive to $\Delta_0$ for a wide range of $\Delta_0$.

\section{Validity of the locally constant field approximation}
\label{sec: LCFA}

The locally constant field approximation (LCFA) mainly relies on the condition that the relevant formation length is much shorter than the scale length at which the background fields vary \cite{Yakimenko2019}. In the quantum regime, the formation length $l_f$ for photon emission in a collision is $l_f\sim \lambdabar_c \gamma \chi_{e \, max}^{-2/3}$ where $\lambdabar_c=\hbar/mc=3.86\times 10^{-11}\ \mathrm{cm}$ is the reduced Compton length. For a beam-beam collision, the scale length of background fields is the beam length $\sigma_{z}$. Therefore, the LCFA is valid when $l_f\ll \sigma_{z}$. For the collision parameters studied in this paper (see Fig. \ref{fig: Sw_Symp_chi200} in Sec. \ref{sec: 3D_PIC_simulation}), $l_f\sim 2\ \mathrm{nm}\ll \sigma_{z}=30.7\ \mathrm{nm}$. Therefore, the LCFA is valid for the collisions considered in our study.

After some algebra, the general condition $l_f\ll \sigma_{z}$ for the LCFA to be valid in a collision can be expressed as 
\begin{equation} 
\left(\frac{\sigma_{0}}{d_e}\right)^{2/3}\frac{\sigma_{z}}{d_e} \gg \left(\gamma \frac{\lambdabar_c}{d_e}\right)^{1/3},
\label{Eq: LCFA}
\end{equation}
where $d_e=c/\omega_{p}$ is the electron skin depth, with $\omega_{p}=\sqrt{4\pi n_0e^2/m}$ the plasma frequency at the peak density $n_0$. For dense beams employed in our study, the beam dimensions ($\sigma_0$ and $\sigma_{z}$) are much larger than $d_e$, and the condition Eq. \eqref{Eq: LCFA} can thus be well satisfied.

\emph{LCFA dependence on the photon energies.} The LCFA also depends on the energy of emitted photons, as studied by \cite{Piazza2018,Fabrizio2019}. The LCFA can be violated for the photon emission at low energies, i.e., $\xi \ll \chi_e/a_0^3$. Here $a_0=\sqrt{\frac{2}{\pi^3}}\frac{r_e}{\sigma_0}N_0$ is the normalized (dimensionless) amplitude of the beam field \cite{Fabrizio2019}. This consideration indicates that the LCFA is valid for photon emission at higher energies, i.e.,
\begin{equation}
\xi ^{{\mathrm{valid \ LCFA}}}\gtrsim \frac{\chi_{e \, max}}{a_0^3}=61 \left( \frac{\sigma_0}{r_e} \right)^3\frac{\chi_{e \, max}}{N_0^3}.
\label{Eq: LCFA_photon_energy_condition}
\end{equation}
For example, in the collision shown by Fig. \ref{fig: Sw_Symp_chi200}, the above equation determines $\xi ^{{\mathrm{valid \ LCFA}}} \gtrsim 0.006$. We note that this restriction of photon energies does not have any noticeable impacts on the results presented in our work because our work is focused on the quantum regime which features a high probability of radiating high-energy photons ($\xi \sim 1 $) as shown by the emitted photon spectrum (Fig. \ref{fig: Sw_Symp_chi200}). Therefore, the photon emission will be fully dominated by high-energy photons where the LCFA well holds. For the collision in Fig. \ref{fig: Sw_Symp_chi200}, the PIC simulation shows that the energy of photons with $\xi \le 0.006$ composes only $0.1\%$ of all the recorded photons.

These estimates, along with results presented in other papers \cite{Fabrizio2019,Yakimenko2019} provide the basis to determine the range of physical parameters to explore within this approximation - this will be further explored in a future publication.

\section{Spectral brightness}
\label{sec: quality_gammaray}

Ultrashort ($\lesssim \mathrm{fs}$) and interpenetrating gamma-ray beams are produced in the collision. Each gamma-ray beam is highly collimated along the beam propagation direction, with a divergence angle of $\theta_0 \sim D\sigma_0/\sigma_z\ll 1$ \cite{Chen1988}. For the beam parameters in Fig. \ref{fig: Sw_Symp_chi200}, $\theta _0\lesssim 1\ \mathrm{mrad}$.

Another important parameter that characterizes the properties of a light source is the brightness. We show that the spectral brightness $B(\xi)$ is $\propto \xi \mathcal{S}_\omega(\xi)$ (see Appendix \ref{sec: appendix_Brightness}). Therefore, $B(\xi)$ also features a sharp peak ($B_{peak}$) at $\xi = \xi_{peak}$. The collision in Fig. \ref{fig: Sw_Symp_chi200} shows $B_{peak}\simeq 6.16\times 10^{30}\ \mathrm{photons}/(\mathrm{s}\ \mathrm{mm}^2\ \mathrm{mrad}^2\ 0.1\% \mathrm{BW})$. Given their unique features, these ultrabright gamma-ray beams, at $100 \, \mathrm{GeV}$-level photon energies ($\mathcal{E}_\gamma \simeq \mathcal{E}_0$), can lead to multiple applications \cite{Asner2003,Gronberg2014,Telnov2018,Takahashi2019,Barklow2022}.

The peak spectral brightness reported above, $B_{peak}$, of the emitted gamma-ray beams in leptonic collisions can be theoretically estimated as 
\begin{eqnarray}
 && \frac{B(\xi=\xi_{peak})}{\mathrm{photons}/(\mathrm{s}\ \mathrm{mm}^2\ \mathrm{mrad}^2\ 0.1\% \mathrm{BW})} \nonumber \\ 
&\simeq & 3.87\times 10^{32}\frac{\gamma}{N_0}\chi_{e \, max} \nonumber \\
&=&1.16\times 10^{26} \frac{\left(\mathcal{E}_0[\mathrm{GeV}]\right)^2}{\sigma_0[0.1 \mu m]  \ \sigma_z[0.1 \mu m]} .
\label{Eq: B_peak}
\end{eqnarray}
Equation \eqref{Eq: B_peak} provides a good estimate for collisions in both $\chi_{e \, max} \sim 1$ \cite{Fabrizio2019} and $\chi_{e \, max}>1$ regimes, as discussed in the detailed derivation given in Appendix \ref{sec: appendix_Brightness}.

Here, we compare the gamma-ray beams shown in our study with other existing or proposed light sources. One of the most well-established methods of producing X- and gamma rays is the laser-Compton scattering light source \cite{Phuoc2012,Gronberg2014,Sarri2014,Yu2016,Takahashi2019}. For this method, the brightness of the resultant light beams can be estimated as
\begin{eqnarray}
    B \sim \frac{1}{4\pi^2}\frac{N_0\alpha\omega_0\gamma^2}{\sigma_0^2a_0}.
    \label{Eq: Brightness_Compton}
\end{eqnarray}
The detailed derivation of the estimate Eq. \eqref{Eq: Brightness_Compton} is given in Appendix \ref{sec: appendix_Brightness}. For an optical laser with $\lambda_0\sim 800 \ \mathrm{nm}$ ($\omega_0 \sim 10^{15}\ rad/s$), colliding with a tightly-focused electron beam $\sigma_0 \sim 0.1 \ \mu m$, we have $B \sim 10^{13}N_0\gamma^2/a_0\ \mathrm{photons}/(\mathrm{s}\ \mathrm{mm}^2\ \mathrm{mrad}^2\ 0.1\% \mathrm{BW})$. For $100\ \mathrm{pC}$ ($N_0 \sim 10^9$), $10$-GeV electron beams and lasers with intensity $a_0 \sim  1$, the quantum parameter $\chi_e \sim 0.01 \ll 1$, and the corresponding maximum brightness could reach $10^{30}\ \mathrm{photons}/(\mathrm{s}\ \mathrm{mm}^2\ \mathrm{mrad}^2\ 0.1\% \mathrm{BW})$, which can be on the same order as the brightness in our study but is radiated at much lower photon frequency (around 100 MeV's).

We also note that the ultrahigh brightness $B_{peak}$ observed here is comparable to the state-of-the-art X-ray sources based on free-electron lasers (FELs) \cite{Tiedtke2009, Boutet2010,Huang2013}. For the configuration of laser-electron collision in the SF-QED regime where $\chi_e\gtrsim 1$ (which is planned to be demonstrated in the near future) \cite{Turner2022}, the brightness of the resultant gamma-ray beams is estimated to be $\sim 10^{24} \ \mathrm{photons}/(\mathrm{s}\ \mathrm{mm}^2\ \mathrm{mrad}^2\ 0.1\% \mathrm{BW}) $, much lower than the gamma-ray beams observed in our study.

\section{Conclusions}
\label{sec: conclusion}

In conclusion, we first derived the analytical spectrum of the integrated beamstrahlung from collisions between round leptonic beams for the quantum regime where $\chi_e \gtrsim 1$. A clear signature for SF-QED as a sharp peak in the spectrum close to the beam energy is identified. We demonstrate that these collisions, with appropriately chosen parameters, preserve the characteristic spectral peak predicted for single-particle radiation dynamics, at a higher $\chi_\mathrm{thr}$, thus providing an excellent platform for precisely probing SF-QED. We then studied the impact of beam-beam effects, misalignment, emittance and energy spread of the colliding beams on beamstrahlung and the characteristic SF-QED signatures. Disruption can significantly affect (or couple with) the beamstrahlung and pair creation, depending on the relation between $\overline{N_{\gamma}}$ and $D$. Ultrashort ($\lesssim \mathrm{fs}$), ultrabright, highly collimated, and interpenetrating gamma-ray beams, at $100 \, \mathrm{GeV}$-level photon energies, are shown to be produced. The theoretical analysis and predictions in this paper have been confirmed by 3D PIC simulations.

\begin{acknowledgments}
	This work was supported by the European Research Council (ERC-2015-AdG grant No. 695088), FCT (Portugal) Grants No. 2022.02230.PTDC (X-MASER), UIDB/FIS/50010/2020 - PESTB 2020-23, Grants No. CEECIND/04050/2021 and PTDC/FIS-PLA/ 3800/2021. W.L.Z. acknowledges the starting fund provided by East China University of Technology. We acknowledge PRACE for awarding us access to MareNostrum at Barcelona Supercomputing Center (BSC, Spain). Simulations were performed at the IST cluster (Lisbon, Portugal) and at MareNostrum.
\end{acknowledgments}

\appendix
\section{Derivation of the integrated beamstrahlung spectrum $\mathcal{S}_\omega(\xi)$ in the quantum regime}
\label{sec: appendix_Analytical_spectrum}

Our analytical model and the setup of collisions under study are introduced in the main text (see Sec. \ref{sec: parameter_setup}). The relevant notations are also defined in the main text. In the theoretical analysis, $z$ is the longitudinal coordinate co-moving with the beam (see Ref. \cite{Chen1988} for details of the associated coordinate system used for collision study). However, in the PIC simulations, shown in Sec. \ref{sec: 3D_PIC_simulation}, $z$ represents the fixed longitudinal coordinates.

The differential probability rate of photon emission in the quantum regime is given by \cite{Ritus1985}
\begin{equation}
\frac{d^2W}{dtd\xi}=\frac{\alpha}{\sqrt{3}\pi \tau_c \gamma }\left[\mathrm{Int}K_{5/3} + \frac{\xi ^2}{1-\xi}K_{2/3}(b) \right].
\label{Eq: Suppl_W_omega}
\end{equation}
Here, $\xi=\mathcal{E}_\gamma/\mathcal{E}_0$ where $\mathcal{E}_\gamma$ is the energy of the radiated photon. $\tau_c = \hbar/mc^2$ is the Compton time. $\alpha$ is the fine-structure constant. $b=2/(3\chi_{e}(r,t))\xi/(1-\xi)$ and $\mathrm{Int}K_{5/3}(b)=\int_{b}^{\infty}K_{5/3}(q)dq$, with $K_{\nu}$ the modified Bessel function of the second kind. 

Equation \eqref{Eq: Suppl_W_omega} shows that the probability of photon emission is solely determined by the local $\chi_e(r,t)$ of the emitting particle. It was known that the spectrum of $d^2W/dtd\xi$ [Eq. \eqref{Eq: Suppl_W_omega}], as a function of $\xi$, presents a peak at $\xi \simeq 1$ for $\chi_{e}>16$ \cite{Esberg2009,Bulanov2013,Tamburini2020}. We further investigated the corresponding power spectrum given by $\mathcal{E}_0\xi d^2W/dtd\xi$. The power spectrum is also found to feature a peak at $\xi =\xi_{peak}\simeq 1$. We define $b_{peak}=2/(3\chi_{e})\xi _{peak}/(1-\xi _{peak})$ to characterize the position of this spectral peak. $b_{peak}$ converges fast to an asymptotic value, i.e., $b_{peak} \rightarrow 0.417$, when $\chi_{e}$ increases. This asymptotic value can be obtained by solving $\partial_b (\xi d^2W/dtd\xi) =0$ in the limit of $\chi_e\gg 1$. If one assumes $b_{peak}\simeq 0.417$, the amplitude of this characteristic peak in the power spectrum is given by
\begin{equation}
\left(\mathcal{E}_0\xi \frac{d^2W}{dtd\xi}\right)_{peak}\simeq 0.17 \frac{\alpha mc^2}{\tau_c}\chi_e .
\label{Eq: Suppl_W_omega_peak}
\end{equation}
This scaling is in excellent agreement with the numerical cross-check. Equation \eqref{Eq: Suppl_W_omega_peak} shows an interesting property of this characteristic peak whose amplitude is solely determined by $\chi_{e}$.
 
The time-integrated photon spectrum emitted from a single particle in a collision is given by $s_\omega (\xi,r)=\int_{-\infty}^{\infty}(d^2W/dtd\xi) dt$ \cite{Fabrizio2019}. The integrated spectrum from the whole beam, normalized by the particle number, is given by
\begin{equation}
\mathcal{S}_\omega(\xi)=\frac{\int_0^{\infty} s_\omega (\xi,r)f(r)rdr}{\int_0^{\infty}f(r)rdr}.
\label{Eq: Suppl_Sw_definition}
\end{equation} 

In order to obtain analytical results for $\mathcal{S}_\omega(\xi)$, we will consider two approximations. \emph{The first approximation} is to approximate the special functions in $d^2W/dtd\xi$ [Eq. \eqref{Eq: Suppl_W_omega}], including $\mathrm{Int}K_{5/3}(b)$ and $K_{2/3}(b)$. \emph{The second approximation} is to approximate the Gaussian beam profile by a uniform profile, which allows to perform the double integral for $\mathcal{S}_\omega(\xi)$.

 \begin{figure*}
	\includegraphics[width=15cm,height=8.46cm]{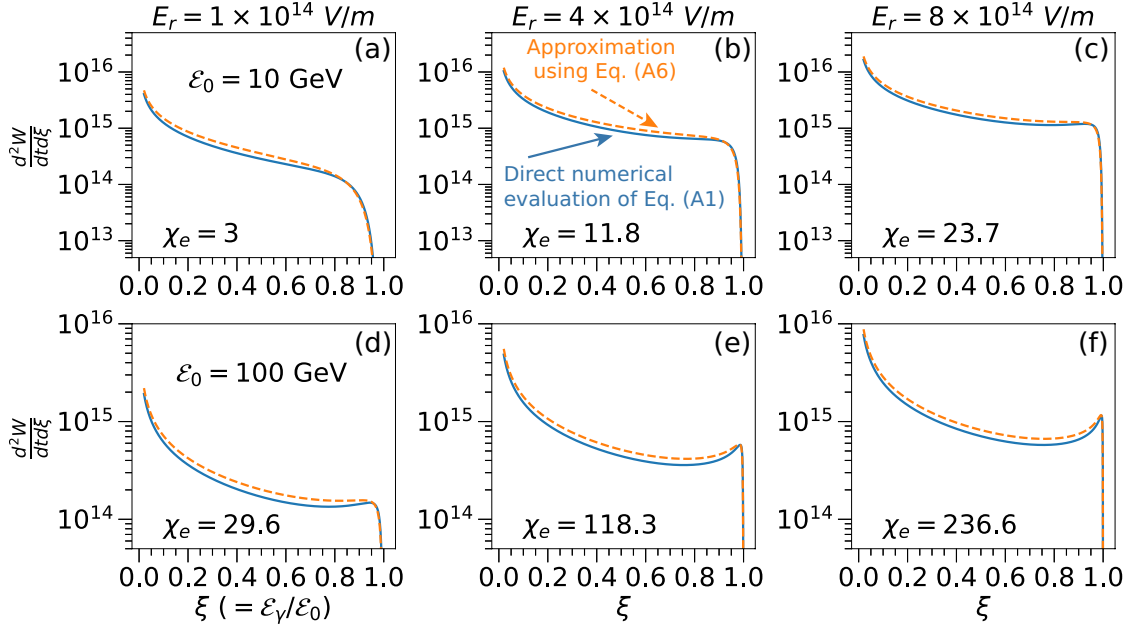}
	\caption{(Color online). Comparison between the differential probability rate of the nonlinear Compton process ($d^2 W/dt d\xi$, given by Eq. \eqref{Eq: Suppl_W_omega} and shown by the blue solid line) and its approximation proposed by this work (using Eq. \eqref{Eq: Suppl_approx_IntK53}, shown by the orange dashed line). The comparison is examined for different electron energy $\mathcal{E}_0$ and field strength (electric field $E_r$ of the beams). The corresponding $\chi_e$ of the electron under study is given by $\chi_e=2(\mathcal{E}_0/mc^2)E_r/E_s$. Upper panel: an electron with $\mathcal{E}_0 = 10 \ \mathrm{GeV}$ under fields of $1\times 10^{14}\ V/m$ in (a), $4\times 10^{14}\ V/m$ in (b), and $8\times 10^{14}\ V/m$ in (c), respectively. Lower panel: an electron with $\mathcal{E}_0 = 100 \ \mathrm{GeV}$ under the same fields with the upper panel.} 
	\label{fig: Suppl_Fig1_com_Ww}
\end{figure*}

\emph{The first approximation.} Both $\mathrm{Int}K_{5/3}(b)$ and $K_{2/3}(b)$ can be well approximated in the $\chi_{e}\gg 1$ regime. In this regime, $b\ll 1$ for most photon emission. We first approximate the $K_{2/3}(b)$ term, since it is dominant in determining high-$\xi$ ($\xi \simeq 1$) photon emission which is of our interest in the quantum regime. The asymptotic analysis of $K_{2/3} (b)$ gives $K_{2/3}(b)\propto  b^{-2/3}$ for $b\ll 1$, and $\propto \exp (-b)$ for $b\gg 1$. One can therefore approximate $K_{2/3}(b)$ as
\begin{equation}
K_{2/3}(b) \simeq k_{2/3}b^{-2/3}\exp (-b)
\label{Eq: Suppl_approx_K23},
\end{equation}
where $k_{2/3}\sim \mathcal{O}(1)$ is the fitting coefficient. Because the $K_{2/3}(b)$ term is critical for high-$\xi$ photon emission, we seek a good approximation of $K_{2/3}(b)$ which matches the characteristic peak of the power spectrum $\mathcal{E}_0\xi d^2W/dtd\xi$ as analyzed in Eq. \eqref{Eq: Suppl_W_omega_peak}. Because the position of this peak obeys $b_{peak}\rightarrow 0.417$ in the $\chi_{e}\gg 1$ regime, we can set the approximation Eq. \eqref{Eq: Suppl_approx_K23} to match the spectral peak, i.e., $K_{2/3}(0.417)= k_{2/3}(0.417)^{-2/3}\exp (-0.417)$, resulting in $k_{2/3}=1.23$.

The approximation of $\mathrm{Int}K_{5/3}$ can be obtained as follows. For $b\gg 1$, $\mathrm{Int}K_{5/3}$ also scales as $\propto \exp (-b)$ \cite{Fabrizio2019}. In addition, one has the following identity for $\mathrm{Int}K_{5/3}(b)$, i.e.,
\begin{equation}
\int_{b}^{\infty}K_{5/3}(q)dq \equiv 2K_{2/3}(b)-\int_{b}^{\infty} K_{1/3}(q) dq
\label{Eq: Suppl_identity_BesselK}.
\end{equation}
Because $K_{2/3}(b)\gg \int_{b}^{\infty} K_{1/3}(q) dq$ for $b\ll 1$, one can have $\mathrm{Int}K_{5/3} = \int_{b}^{\infty}K_{5/3}(q)dq \simeq 2K_{2/3}(b)$ for $b\ll 1$. 

The above analysis indicates that $\mathrm{Int}K_{5/3}$ can be approximated as
\begin{equation}
\mathrm{Int}K_{5/3}\simeq 2K_{2/3}(b)\simeq 2k_{2/3}b^{-2/3}\exp (-b).
\label{Eq: Suppl_approx_IntK53}
\end{equation}
Using Eq. \eqref{Eq: Suppl_approx_IntK53}, one can have an excellent approximation of $d^2W/dtd\xi$ [Eq. \eqref{Eq: Suppl_W_omega}]. This has been verified as shown in Fig. \ref{fig: Suppl_Fig1_com_Ww}. Our approximation (orange dashed line) agrees well with the direct numerical evaluation of Eq. \eqref{Eq: Suppl_W_omega} (blue solid line), for the full range of $\xi \in (0,1)$. Particularly, the approximation accurately characterizes the high-energy photon emission (with $\xi\sim 1$). This is due to the fact that we had sought a good approximation of $d^2 W/dt d\xi$ which can match the characteristic peak of the power spectrum $\xi d^2 W/dt d\xi$. In addition, our approximation works well for a broad quantum regime where $\chi_e\gtrsim 1$. This lays the foundation that our analytical beamstrahlung spectrum (to be derived later based on this approximation) can give the accurate photon emission in both the $\chi_{e \, max}\sim 1$ and the $\chi_{e \, max}\gg 1$ regimes. 
 
\emph{The second approximation.} The Gaussian profile of the colliding beams can be approximated by a uniform cylinder of radius $\sigma _{0A}$ and length $\sigma_{zA}=2\sqrt{2}\sigma_z$, with a constant density $n_{0A}$. The profile of the approximated cylinder is given by $f_A(r)\equiv1$ for $r\le \sigma _{0A}$, and $g_A(z)\equiv 1$ for $z_c-\sigma_{zA}/2 \le  z \le z_c+\sigma_{zA}/2$. We further impose the requirement that the approximated uniform profile should match $\chi_{e \, max}$ of the original Gaussian profile. The electric field of the approximated uniform beam is given by $E_{r A}=2\pi en_{0A}r$. The maximum $E_{r A}$, given by $E_{r A, \, max}=2\pi en_{0A}\sigma _{0A}$, should match the peak field of the original Gaussian beam, leading to $n_{0A}\sigma_{0A}=0.9n_0\sigma_0$. We have justified that a collision with the approximated uniform profile can preserve the $dN/d\chi_e$ distribution of particles in the collision between the original Gaussian beams, as shown by Fig. \ref{fig: Gaus_Unif_justification} in Sec. \ref{sec: analytical_spectrum}.

\emph{The integrated radiation spectrum $\mathcal{S}_\omega(\xi)$.} As analyzed in \emph{the second approximation}, the radiation from a collision between round Gaussian beams can be approximated by that from a collision between the approximated uniform cylinder beams. With the approximated uniform profile, the corresponding $\chi_{e}$ is given by $\chi_{eA}=2\gamma E_{r A}/E_s$, and the $b$ parameter in $d^2W/dtd\xi$ [Eq. \eqref{Eq: Suppl_W_omega}] is given by
\begin{equation}
b=\frac{2}{3\chi_{eA}(r,t)}\frac{\xi}{1-\xi}=\frac{2}{3\chi_{e \, max}}\frac{\xi}{1-\xi}\frac{\sigma_{0A}}{r}=b_0\frac{\sigma_{0A}}{r},
\label{Eq: Suppl_b}
\end{equation}
where $b_0=2/(3\chi_{e \, max})\xi/(1-\xi)$.

The normalized integrated spectrum $\mathcal{S}_\omega(\xi)$ is therefore calculated as
\begin{eqnarray}
&&\mathcal{S}_\omega^{Gaussian}(\xi) =\int _0^{\infty} s_\omega (\xi,r)f(r)rdr/\int _0^{\infty} f(r)rdr \nonumber \\
&& \simeq \mathcal{S}_\omega^{Approximated}(\xi) \nonumber \\
&&=\int _0^{\sigma_{0A}} s_{\omega A} (\xi,r)f_A(r)rdr/\int _0^{\sigma_{0A}} f_A(r)rdr \nonumber \\
&& =\int _0^{\sigma_{0A}} \left( \int _0^{\sigma_{zA}/2c} \frac{d^2W}{dtd\xi} dt \right) f_A(r)rdr / \int  _0^{\sigma_{0A}} f_A(r)rdr \nonumber \\
&&=\frac{2\sqrt{2}\alpha \sigma_{z}}{\sqrt{3}\pi\tau_c\gamma c} k_{2/3} \left(2+\frac{\xi ^2}{1-\xi} \right)b_0^2\ \Gamma\left[-\frac{8}{3}, b_0\right],
\label{Eq: Suppl_Sw}
\end{eqnarray}
where $\Gamma$ is the incomplete Gamma function \cite{IncGamma_Book,IncGamma_Online}. Equation \eqref{Eq: Suppl_Sw} is shown to give the accurate photon emission in both the $\chi_{e \, max}\sim 1$ and the $\chi_{e \, max}> 1$ regimes. This is confirmed by the numerical calculation of $\mathcal{S}_\omega(\xi)$ without employing approximations, i.e., using the exact $d^2W/dtd\xi$ in Eq. \eqref{Eq: Suppl_W_omega} and original Gaussian beam profiles. In addition, Equation \eqref{Eq: Suppl_Sw} is also verified by 3-dimensional (3D) QED particle-in-cell (PIC) simulations as shown in the main text.

\section{Estimate of the spectral brightness $B(\xi)$ of the emitted gamma-rays}
\label{sec: appendix_Brightness}

The spectral brightness of a photon beam is defined as \cite{Wiedemann2007}
\begin{equation}
B(\xi)=\frac{\frac{dN_\gamma}{dt}}{4\pi ^2\sigma_{x}\sigma_{y}\sigma_{x'}\sigma_{y'}\frac{\Delta \omega}{\omega}},
\label{Eq: Suppl_definition_B}
\end{equation}
where $dN_\gamma/dt$ is the photon flux (per second) at a specific photon frequency $\omega$ with a bandwidth $\Delta \omega/\omega$. Here, $\Delta \omega/\omega=\Delta \xi/\xi$, as $\xi=\mathcal{E}_\gamma /\mathcal{E}_0=\hbar \omega/\mathcal{E}_0$. In the community of synchrotron light source, the bandwidth is usually chosen to be $\Delta \xi/\xi=0.1\%$. $\sigma_{x}$ and $\sigma_{y}$ are root mean square (RMS) values of transverse sizes of the photon beam in the $x$ and $y$ directions, respectively. $\sigma_{x'}$ and $\sigma_{y'}$ are the corresponding RMS values of solid angles (momentum divergence of the photon beam). 

The number of photons at energy $\xi$ with a specific bandwidth $\Delta \xi/\xi$ is given by
\begin{equation}
{N}_{\gamma} (\xi) = N_0 \mathcal{S}_\omega(\xi)\Delta \xi = N_0 \xi \mathcal{S}_\omega(\xi) \frac{\Delta \xi}{\xi}.
\label{Eq: Suppl_N_gamma}
\end{equation}
This relation shows that ${N}_{\gamma} (\xi) \propto \xi \mathcal{S}_\omega(\xi)$ when $\Delta \xi/\xi$ is fixed. Therefore, one has $B(\xi)\propto \xi \mathcal{S}_\omega(\xi)$, assuming that photons at different energy $\xi$ have the same pulse duration. Since $B(\xi)\propto \xi \mathcal{S}_\omega(\xi)$, the spectral brightness will also feature a a sharp peak ($B_{peak}$) at $\xi = \xi_{peak}\simeq 1$, same with the $\xi \mathcal{S}_\omega$ spectrum. Using Eq. \eqref{Eq: Suppl_N_gamma}, the number of photons at energies $[\xi_{peak}, \xi_{peak}+\Delta \xi]$ is simply given by
\begin{equation}
N_{\gamma} (\xi_{peak})=N_0P_{\xi \mathcal{S}_\omega} \frac{\Delta \xi _{peak}}{\xi_{peak}},
\label{Eq: Suppl_N_gamma_xi_peak}
\end{equation}
where $P_{\xi \mathcal{S}_\omega}$ is the amplitude of the peak in the $\xi \mathcal{S}_\omega$ spectrum, given by Eq. \eqref{Eq: P_xiSw_Sym} in the main text. 

The emitted photon beam has the same emittance, including the spatial size and divergence angles, with the emitting beam in a collision. This allows us to conveniently set $\sigma_{x}=\sigma_{y}\simeq \sigma_{0}$ and $\tau_\gamma \simeq 2\sqrt{2}\sigma_{z}/c$, where $\tau_\gamma $ is the pulse duration of the photon beam. The divergence angle of the photon beam can be approximated as the deflection angle of the beam particles, due to the disruption effect caused by the fields from the oncoming beam \cite{Chen1988}. The deflection angle of a particle initially at $r=r_0$ is calculated as $\theta_0(r_0)=r_e N_0r_0/(\gamma \sigma_{0}^2)$ \cite{Chen1988}. For the particles at $r_0 \simeq r_{peak}$ where $r_{peak}$ depicts the position of the peak fields of the oncoming beam, they will undergo fastest and strongest radiation and deflection. For a Gaussian beam, $r_{peak}=1.6\sigma_0$. We therefore specifically employ the deflection angle of particles at $r_0\simeq r_{peak}$ to estimate the average divergence angle of the whole radiated photon beam, i.e.,
\begin{equation}
\sigma_{x'}=\sigma_{y'}\simeq \theta_0 (r\simeq r_{peak})\simeq \frac{r_eN_0r_{peak}}{\gamma \sigma_0^2}.
\label{Eq: Suppl_sigma_xy_pri}
\end{equation}
This divergence estimation Eq. \eqref{Eq: Suppl_sigma_xy_pri} agrees well with the 3D PIC simulations.

With the above analysis, the peak spectral brightness at $\xi = \xi_{peak}$ is estimated as
\begin{eqnarray}
&& \frac{B_{peak}(\xi_{peak})}{\mathrm{photons}/(\mathrm{s}\ \mathrm{mm}^2\ \mathrm{mrad}^2\ 0.1\% \mathrm{BW})}\nonumber \\
&= & \frac{N_\gamma (\xi_{peak})/\tau_\gamma}{4\pi ^2\sigma_{x}\sigma_{y}\sigma_{x'}\sigma_{y'}\Delta \xi/\xi_{peak}}  \nonumber \\ 
&\simeq & 3.87\times 10^{32}\frac{\gamma}{N_0}\chi_{e \, max} \nonumber \\
&=&1.16\times 10^{26} \frac{\left(\mathcal{E}_0[\mathrm{GeV}]\right)^2}{\sigma_0[0.1 \mu m]  \ \sigma_z[0.1 \mu m]}.
\label{Eq: Suppl_B_peak}
\end{eqnarray}

Equation \eqref{Eq: Suppl_B_peak} provides a good estimation for the peak brightness of beamstrahlung photon beams in both the $\chi_{e \, max} \sim 1$ and the $\chi_{e \, max}>1$ regimes. This is confirmed by 3D PIC simulations.

\emph{Estimate of the brightness in laser-Compton scattering and synchrotron light sources.}

The brightness for the Compton/synchrotron light sources in the classical regime can be estimated as follows. The photon flux $dN_{\gamma}/dt$ around the critical frequency $\omega_c$ can be estimated using the classical Compton/synchrotron radiated power $P$
\begin{equation}
    \frac{dN_{\gamma}}{dt} \sim N_0\frac{P}{\hbar\omega_c} \sim N_0\frac{\alpha (\chi_e mc^2)^2/\hbar}{\chi_e \gamma mc^2} \sim N_0\alpha \omega_0 a_0,
    \label{Eq: Suppl_dNg_dt_Compton}
\end{equation}
with $\hbar\omega_c \sim \chi_e \gamma mc^2$ and $\chi_e \ll 1$. On the other hand, the divergence angle is approximated by \cite{Vranic2016}
\begin{equation}
    \theta_0 \sim \frac{a_0}{\gamma},
    \label{Eq: Suppl_theta_Compton}
\end{equation}
which is, surprisingly, the same as the result obtained in Eq. \eqref{Eq: Suppl_sigma_xy_pri} if one substitutes $a_0$ here (in Eq. \eqref{Eq: Suppl_theta_Compton}) by the normalized beam field $a_0 \simeq r_eN_0/\sigma_0$ (defined in Sec. \ref{sec: LCFA}).

With the above analysis, we can estimate the corresponding brightness
\begin{eqnarray}
    B \sim \frac{1}{4\pi^2}\frac{N_0\alpha\omega_0\gamma^2}{\sigma_0^2a_0}.
\end{eqnarray}
For an optical laser with $\lambda_0\sim 800 \ \mathrm{nm}$ ($\omega_0 \sim 10^{15}\ rad/s$), colliding with a tightly-focused electron beam $\sigma_0 \sim 0.1 \ \mu m$, we have $B \sim 10^{13}N_0\gamma^2/a_0\ \mathrm{photons}/(\mathrm{s}\ \mathrm{mm}^2\ \mathrm{mrad}^2\ 0.1\% \mathrm{BW})$. For $100\ \mathrm{pC}$ ($N_0 \sim 10^9$), $10$-GeV electron beams and lasers with intensity $a_0 \sim  1$, the quantum parameter $\chi_e \sim 0.01 \ll 1$, and the corresponding maximum brightness could reach $10^{30}\ \mathrm{photons}/(\mathrm{s}\ \mathrm{mm}^2\ \mathrm{mrad}^2\ 0.1\% \mathrm{BW})$, which can be on the same order as the brightness in our study but is radiated at much lower photon frequency (around 100 MeV's).

\newpage


\begin{thebibliography}{99}
    \bibitem{Shiltsev2021}V. Shiltsev and F. Zimmermann, Rev. Mod. Phys. 93, 015006 (2021).
    \bibitem{Gray2021}H. M. Gray, Reviews in Physics 6, 100053 (2021).
    \bibitem{ILC_ExecutiveSummary}T. Behnke, et al., The International Linear Collider Technical Design Report - Volume 1: Executive Summar, arXiv:1306.6327 (2013).  
    \bibitem{CLIC_Implementation2017}M. Aicheler, P. N. Burrows, N. Catalan, R. Corsini, M. Draper, J. Osborne, D. Schulte, S. Stapnes, and M. Stuart, 2019, Eds., CERN Yellow Report No. CERN-2018-010-M.
    \bibitem{FACETII_report}FACET-II Technical Design Report No. SLAC-R-1072 (2016).
    \bibitem{EurStrParPhys2022}European Strategy for Particle Physics - Accelerator R$\&$D Roadmap, N. Mounet (ed.), CERN Yellow Reports: Monographs, CERN-2022-001 (CERN, Geneva, 2022).

    \bibitem{Chen1992}P. Chen, Phys. Rev. D 46, 1186 (1992).
    \bibitem{Yokoya1992}K. Yokoya, P. Chen, "Beam-beam phenomena in linear colliders." Frontiers of Particle Beams: Intensity Limitations. Lecture Notes in Physics, vol. 400, pp. 415, 1992, Springer, Berlin, Heidelberg.
    \bibitem{Schulte2017}D. Schulte, Beam-Beam Effects in Linear Colliders, in CERN yellow reports: School Proceedings, vol. 3, pp. 431–447. ISSN 2519-805X (2017).
      
    \bibitem{Barklow2023}Tim Barklow, et al., JINST 18, P09022 (2023).

    \bibitem{Ritus1985}V. I. Ritus, Quantum effects of the interaction of elementary particles with an intense electromagnetic field, J. Sov. Laser Res. 6, 497 (1985).
    \bibitem{Baier1998}V. N. Baier, V. M. Katkov, and V. M. Strakhovenko, Electromagnetic Processes at High Energies in Oriented Single Crystals (World Scientific, Singapore, 1998).
    \bibitem{Piazza2012}A. Di Piazza, C. Müller, K. Z. Hatsagortsyan, and C. H. Keitel, Rev. Mod. Phys. 84, 1177 (2012).
    \bibitem{Bulanov2013}S. S. Bulanov, C. B. Schroeder, E. Esarey, and W. P. Leemans, Phys. Rev. A 87, 062110 (2013).
    \bibitem{Gonoskov2022}A. Gonoskov, et al., Rev. Mod. Phys. 94, 045001 (2022).
    \bibitem{Fedotov2022}A. Fedotov, et al., Physics Reports, 1010, 1–138 (2023).
        
    \bibitem{Schroeder2010}C. B. Schroeder, et al., Phys. Rev. ST Accel. Beams 13, 101301 (2010).
    \bibitem{ALEGRO2019}ALEGRO collaboration, Towards an Advanced Linear International Collider, arXiv:1901.10370 (2019).
    \bibitem{Clark2022}C. Clarke, et al., JINST 17, T05009 (2022).
    \bibitem{Adli2022}E. Adli, JINST, 17, T05006 (2022).
    \bibitem{England2022}R. J. England, et al., JINST, 17, P05012 (2022).
    \bibitem{Snowmass21TaskForce}Thomas Roser, et al., JINST 18, P05018 (2023).

    \bibitem{Noble1987}R. J. Noble, Nuclear Instruments and Methods in Physics Research Section A, 256, 427 (1987).
    \bibitem{Chen1989}P. Chen and V. I. Telnov, Phys. Rev. Lett. 63, 1796 (1989).
    \bibitem{Fabrizio2019}F. Del Gaudio, et al., Physical Review Accelerators and Beams, 22, 023402 (2019).
    \bibitem{Yakimenko2019}V. Yakimenko, et al., Phys. Rev. Lett. 122, 190404 (2019).      
    \bibitem{Tamburini2020}M. Tamburini and S. Meuren, Phys. Rev. D 104, L091903 (2021).
    \bibitem{Samsonov2021}A. S. Samsonov, et al., New J. Phys. 23, 103040 (2021).

    \bibitem{Burke1997}D. L. Burke, et al., Phys. Rev. Lett. 79, 1626 (1997).
    \bibitem{Blackburn2014}T. G. Blackburn, et al., Phys. Rev. Lett. 112, 015001 (2014).
    \bibitem{Piazza2016}A. Di Piazza, Phys. Rev. Lett. 117, 213201 (2016).
    \bibitem{Vranic2016}M. Vranic, et al., New J. Phys. 18, 073035 (2016).
    \bibitem{Lobet2017}M. Lobet, et al., Phys. Rev. Accel. Beams 20, 043401 (2017).
    \bibitem{Gales2018}S. Gales, et al., Rep. Prog. Phys. 81, 094301 (2018).
    \bibitem{Poder2018}K. Poder, et al., Phys. Rev. X 8, 031004 (2018).
    \bibitem{Niel2018b}F. Niel, et al., Plasma Phys. Control. Fusion 60, 094002 (2018).
    \bibitem{Cole2018}J. M. Cole, et al., Phys. Rev X 8, 011020 (2018).
    \bibitem{Baumann2019b}C. Baumann, et al., Sci. Rep. 9, 9407 (2019).
    \bibitem{Albert2020}F. Albert, et al., New J. Phys. 23, 031101 (2021).
    \bibitem{Hu2020}G. Hu, et al., Phys. Rev. A 102, 042218 (2020).
    \bibitem{Fedeli2021}L. Fedeli, et al., Phys. Rev. Lett. 127, 114801 (2021).
    \bibitem{Grech2021}J. P. Zou, et al., High Power Laser Science and Engineering, 3, E2 (2015). 
    \bibitem{Qu2021}Kenan Qu, Sebastian Meuren, and Nathaniel J. Fisch, Phys. Rev. Lett. 127, 095001 (2021).
    \bibitem{Golub2022}A. Golub, S. Villalba-Chávez, and C. Müller, Phys. Rev. D 105, 116016 (2022).
    \bibitem{Turner2022}M. Turner, et al., Eur. Phys. J. D 76, 205 (2022).
    \bibitem{Ahmadiniaz2022}N. Ahmadiniaz, et al., Phys. Rev. D 108, 076005 (2023).
    
    \bibitem{Piazza2020}A. Di Piazza, et al., Phys. Rev. Lett. 124, 044801 (2020).
    
    \bibitem{Matheron2022}A. Matheron, P. San Miguel Claveria, R. Ariniello, et al., Commun. Phys. 6, 141 (2023).
    
    \bibitem{Ritus1972}V. I. Ritus, Radiative corrections in quantum electrodynamics with intense field and their analytical properties, Ann. Phys. (N.Y.) 69, 555 (1972).   
    \bibitem{Fedotov2017}A. M. Fedotov, J. Phys. Conf. Ser. 826, 012027 (2017).
    \bibitem{Podszus2019}T. Podszus and A. Di Piazza, Phys. Rev. D 99, 076004 (2019).
    \bibitem{Ilderton2019_NPQED}A. Ilderton, Phys. Rev. D 99, 085002 (2019).
    \bibitem{Mironov2020}A. A. Mironov, S. Meuren, and A. M. Fedotov, Phys. Rev. D 102, 053005 (2020).
    \bibitem{Heinzl2021}T. Heinzl, A. Ilderton, and B. King, Phys. Rev. Lett. 127, 061601 (2021).

    \bibitem{Chen1988}P. Chen and K. Yokoya, Phys. Rev. D, 38, 987 (1988).

    \bibitem{Piazza2018}A. Di Piazza, et al., Phys. Rev. A 98, 012134 (2018).
    \bibitem{Piazza2019}A. Di Piazza, M. Tamburini, S. Meuren, and C. H. Keitel, Phys. Rev. A 99, 022125 (2019).
    \bibitem{Ilderton_PRA_LCFA_2019}A. Ilderton, B. King, and D. Seipt, Phys. Rev. A 99, 042121 (2019).
    \bibitem{King2020}B. King, Phys. Rev. A 101, 042508 (2020).
    \bibitem{Lv2021}Q. Z. Lv, et al., Phys. Rev. Research 3, 013214 (2021).    

    \bibitem{IncGamma_Book}Incomplete Gamma and Related Functions, Chapter 8, NIST Handbook of Mathematical Functions, Edited by Frank W. J. Olver, Daniel W. Lozier, Ronald F. Boisvert and Charles W. Clark. Cambridge University Press, 2010.
    \bibitem{IncGamma_Online}R. B. Paris, Incomplete Gamma and Related Functions, NIST Digital Library of Mathematical Functions, https://dlmf.nist.gov/8

    \bibitem{Fonseca2002}R. A. Fonseca, L. O. Silva, F. S. Tsung, V. K. Decyk, W. Lu, C. Ren, W. B. Mori, S. Deng, S. Lee, T. Katsouleas, and J. C. Adam, in Computational Science ICCS 2002, Lecture Notes in Computer Science (Springer, Berlin, Heidelberg, 2002), pp. 342–351.    
    \bibitem{Grismayer2016}T. Grismayer, et al., Phys. Plasmas 23, 056706 (2016).
    \bibitem{Zhang2021}W. L. Zhang, T. Grismayer, K. M. Schoeffler, R. A. Fonseca, and L. O. Silva, Phys. Rev. E 103, 013206 (2021).
    \bibitem{Grismayer2021}T. Grismayer, et al., New J. Phys. 23, 095005 (2021).
    
    \bibitem{Esberg2009}J. Esberg and U. I. Uggerhøj, J. Phys. 198, 012007 (2009).

    \bibitem{Martins2010}S. F. Martins, R. A. Fonseca, W. Lu, et al., Nature Phys. 6, 311 (2010).

    \bibitem{Schoeffler2019}K. M. Schoeffler, et al., ApJ 870, 49 (2019).

    \bibitem{ILC_TDR}Chris Adolphsen, et al., The International Linear Collider Technical Design Report - Volume 3.I: Accelerator R$\&$D in the Technical Design Phase, arXiv:1306.6353 (2013).

    \bibitem{Asner2003}D. Asner, et al., Eur. Phys. J. C. 28, 27 (2003).
    \bibitem{Gronberg2014}Jeffrey Gronberg, Rev. Accel. Sci. Techol. 7, 161 (2014).
    \bibitem{Telnov2018}V. I. Telnov, JINST, 13 P03020 (2018).
    \bibitem{Takahashi2019}T. Takahashi, Rev. Accel. Sci. Techol. 10, 215 (2019).
    \bibitem{Barklow2022}T. Barklow, et al., XCC: An X-ray FEL-based $\gamma \gamma$ Collider Higgs Factory, arXiv:2203.08484v2 (2022).

    \bibitem{Phuoc2012}K. Ta Phuoc, et al., Nature Photonics, 6, 308 (2012).
    \bibitem{Sarri2014}G. Sarri, et al., Phys. Rev. Lett. 113, 224801 (2014).
    \bibitem{Yu2016}Changhai Yu, et al., Sci. Rep. 6, 29518 (2016).
    
    \bibitem{Tiedtke2009}K. Tiedtke, et al., New J. Phys. 11, 023029 (2009).
    \bibitem{Boutet2010}S. Boutet and G. J. Williams, New J. Phys. 12, 035024 (2010).
    \bibitem{Huang2013}Z. Huang, Brightness and coherence of synchrotron radiation and FELs. Tech. Rep., SLAC National Accelerator Lab., Menlo Park, CA (United States) (2013).
    
    \bibitem{Wiedemann2007}H. Wiedemann, Particle accelerator physics (3rd ed.), Springer Berlin, Heidelberg, 2007.

\end{thebibliography}
\end{document}